\documentstyle[twoside,fleqn,espcrc2]{article}

\newcommand{\AmS}{{\protect\the\textfont2
  A\kern-.1667em\lower.5ex\hbox{M}\kern-.125emS}}

\title{String Dualities from Matrix Theory: A Summary}

\author{Micha Berkooz\address{School of Natural Sciences,\\
	Institute for Advanced Study \\
        Olden Lane, Princeton, NJ 08540, USA.}
	\thanks{Talk presented at the Strings 97 conference. 
	IASSNS-HEP-97/112}
}

\begin{document}

\begin{abstract}

I review the appearance, within Matrix theory, of the $SL(5,Z)$ U-duality
 group of M-theory on $T^4$, and the duality between M-theory on K3 and 
the Heterotic string on $T^3$. In both cases the duality is geometrical 
and manifest.    

\end{abstract}

\maketitle

\def\ijmp#1#2#3{{\it Int. J. Mod. Phys.} {\bf A#1,} #2 (19#3)}
\def\mpla#1#2#3{{\it Mod.~Phys.~Lett.} {\bf A#1,} #2 (19#3)}
\def\npb#1#2#3{{\it Nucl. Phys.} {\bf B#1,} #2 (19#3)}
\def\np#1#2#3{{\it Nucl. Phys.} {\bf #1,} #2 (19#3)}
\def\plb#1#2#3{{\it Phys. Lett.} {\bf B#1,} #2 (19#3)}
\def\prc#1#2#3{{\it Phys. Rev.} {\bf C#1,} #2 (19#3)}
\def\prd#1#2#3{{\it Phys. Rev.} {\bf D#1,} #2 (19#3)}
\def\pr#1#2#3{{\it Phys. Rev.} {\bf #1,} #2 (19#3)}
\def\prep#1#2#3{{\it Phys. Rep.} {\bf C#1,} #2 (19#3)}
\def\prl#1#2#3{{\it Phys. Rev. Lett.} {\bf #1,} #2 (19#3)}
\def\rmp#1#2#3{{\it Rev. Mod. Phys.} {\bf #1,} #2 (19#3)}

\def\IZ{\relax\ifmmode\mathchoice
{\hbox{\cmss Z\kern-.4em Z}}{\hbox{\cmss Z\kern-.4em Z}}
{\lower.9pt\hbox{\cmsss Z\kern-.4em Z}}
{\lower1.2pt\hbox{\cmsss Z\kern-.4em Z}}\else{\cmss Z\kern-.4em
Z}\fi}

\section{Introduction}  

In the last year a significant amount of evidence has
accumulated in support of the conjecture of Banks, Fischler, Shenker
and Susskind on the non-perturbative formulation of
M-theory\cite{bfss}. 
The conjecture, however, needs to be extended
when describing M-theory compactified down to seven or less non-compact 
spacetime dimensions. 
The details of the extension for M-theory on a 4-torus were discussed 
in \cite{moshe,brs}, on a 5-torus in \cite{brs,nnati}, on $T^5/Z_2$ 
in \cite{nnati}, and a proposal for M-theory on $K3$ was put forward in 
\cite{mkthree}. For a more systematic approach, see \cite{natinew}. 
In this lecture I will discuss some aspects of how the various 
perturbative and non-perturbative symmetries of string theory manifest 
themselves in Matrix theory when non-compact space-time is seven 
dimensional\footnote{The emergence of the U-duality group of M-theory 
on $T^3$ from non-perturbative dynamics in $N=4$ 3+1 SYM is discussed 
in \cite{lennya,origa}. This U-duality group is, however, a subgroup 
of the U-duality of, say, M-theory on $T^4$ with the advantage that in 
the latter it is geometric and manifest, rather than non-perturbative.}.

The focus of these lectures will be the $SL(5,Z)$ duality of M-theory 
on $T^4$ and the duality between M theory compactified on K3 and the 
Heterotic string on $T^3$. The results presented in this lecture were 
obtained in collaboration with M. Rozali and N. Seiberg. I will 
mainly follow the approach in \cite{brs,mkthree} and the reader is 
referred to additional details there. Some of the results 
in section 3 were derived also by \cite{horavb,govin}. The constructions 
presented in this lecture are also similar to those in \cite{dvv}, but 
in the context of Matrix theory.

Additional insight and details can be found in the lectures of N. 
Seiberg and E. Verlinde at this conference.

\section{U-duality of M-theory on $T^4$}

In dimension $d<4$, the Matrix description of M-theory on $T^d$ is 
given by Super-Yang-Mills (SYM) on a ``Dual'' manifold (with 16 Susy) 
\cite{bfss}. This dual manifold is also a d-dimensional torus but the 
radii of the torus are different [1,7,8,12-17], and we will denote it 
by ${\tilde T}^d$. In these dimensions this proposal does indeed 
constitute a complete proposal as these SYM are renormalizeable and 
therefore exist without the need to add degrees of freedom. This 
conjecture, however, has to be extended when describing M-theory 
compactified\footnote{A recent systematic approach is described 
in \cite{natinew,sen}} on $T^4$ (and other 4 or higher dimensional 
manifolds). 
The reason is that the ``SYM on a dual torus'' guarantees that the 
extreme IR of the base-space theory will be roughly correct, but does 
not provide any information on the UV of the theory. Indeed, 4+1 SYM 
is not renormalizeable and one needs to specify its UV in order to make 
sense of such a proposal. The relevant extension was discussed in 
\cite{moshe,brs}
 where it was suggested that M-theory on a
4-torus is given by the large N limit of the (2,0) supersymmetric
field theory compactified on a 5-torus, which we will denote by 
${\tilde T}^5$.

The details of the extension are the following. Let us begin with the 
4+1 SYM picture. For simplicity, the 
torus will be taken to be rectangular. M-theory has 5 dimensionful 
parameters which are the eleven dimensional Planck scale $l_p$ the 4 
periodicities of the torus $L_i$.  
The parameters of the gauge theory 
are the dual torus lengths
$\tilde L_i$ and the gauge coupling $g^2$. They are given by
\cite{moshe,fhrs}:
\begin{equation}
\begin{array}{rcl}
&\tilde L_i= \frac{(2\pi)^2 {l_p}^3}{L_i R} \\
&g^2= \frac{ (2\pi)^6 {l_p}^6}{L_1 L_2 L_3 L_4 R} \\
\end{array}
\label{eq:sigmadef}
\end{equation}
where $R$ is the radius of the compactified light cone.

This gauge theory is not renormalizable and therefore not well defined.
However, it can be used as a low energy effective theory which is valid
at energies below ${1 \over g^2}$ (In this regime several tests of this 
theory as a formulation of M-theory were shown to be successful
\cite{moshe,fhrs}). In order to define our 4+1 dimensional 
nonrenormalizable theory we need to give
more information about its short distance degrees of freedom. In our 
case, we are guided by the
$SL(5,Z)$ U-duality group to suggest that the desired definition of this
theory is in terms of the $(2,0)$ supersymmetric fixed point in six
dimensions.

The $(2,0)$ theory is an interacting quantum field theory at a fixed
point of the renormalization group (for a review, see \cite{sixteen}).
It was first discussed in the context of type IIB compactification on a
singular K3 \cite{wittentwoz}
and later in the context of $N$ nearby 5-branes in M-theory 
\cite{andy,wittenbranes}
 Here we use the same field theory, compactified on ${\tilde T^5}$, as a
definition of M-theory\footnote{Even though the (2,0) theory appears as a 
subsector of M-theory, the way we employ it to define a Matrix model 
for M-theory is not tautological. The (2,0) theory, as a field theory, 
makes perfect sense on its own. We will at most use well established 
 low-energy properties of M-theory to learn about the (2,0) field theory.
}.

We thus propose that M-theory on $T^4$ with radii $L_{1,2,3,4}$ is
described by the large $N$ limit of the (2,0) theory
compactified on a five torus $\tilde T^5$ (with an $SL(5,Z)$ invariant 
choice of spin structure \cite{witten}).
Its five sizes are related to $L_{1,2,3,4}$ in the
following way.  In equation (\ref{eq:sigmadef}) the sizes of the four 
torus and the
coupling of the 4+1 SYM were given. This SYM has an additional
conserved $U(1)$ current given by $j={}^*(F\wedge F)$. This symmetry
is to be identified with the Kaluza-Klein $U(1)$ symmetry of rotating
around the small circle. This determines the circumference of the circle 
to be $g^2\over
2\pi$.  In the 4+1 SYM an $n$ instantons
configuration, which has energy $4\pi^2n\over g^2$, corresponds to a
state with total momentum $n$ around the small circle and hence we
identify its periodicity with ${g^2\over 2\pi}$.  The 4+1 SYM
description only identifies the charge of this $U(1)$ symmetry.

To summarize, we propose that the (2,0) theory is compactified on
$\tilde T^5$ whose sizes are given by
\begin{equation}
\begin{array}{rcl}
&\tilde L_i= \frac{(2\pi)^2 {l_p}^3}{L_i R} \\
&\tilde L_5= \frac{ (2\pi)^5 {l_p}^6}{L_1 L_2 L_3 L_4 R}.\\
\end{array}
\end{equation}

Our main focus is the manifestation of the U-duality group for this 
configuration, but let us briefly discuss some tests of this proposal.
As a first test of this proposal we can restate what we had explained 
before. 
At energies much
smaller than ${1 \over g^2} = {1 \over \tilde L_5}$ and for $\tilde L_5
\ll \tilde L_{1,2,3,4}$ our theory
becomes the 4+1 dimensional SYM. This corresponds to the space-time 
4-torus being much larger than the Planck-scale. The gap to the states 
that carry momentum around $\tilde L_5$ is proportional to the volume 
of the space-time torus. A process of scattering gravitons in a region 
smaller compared to the 4-torus of space-time will yield the correct 
results, up to terms that are suppressed by the volume of this 4-torus. 
In particular general relativity will be a correct local description. 
As another test one can describe the various (particle like) solitons 
in the theory as fluxes in the (2,0) theory.

The most striking feature of this proposal \cite{moshe} is that
this definition of M-theory on $T^4$ makes the U-duality manifest.
The U-duality group in this case is $SL(5,Z)$. It is simply the geometric
duality group of $\tilde T^5$.  This symmetry involves mixing the five
radii $\tilde L$ in a way which is complicated as an action on the
individual $L_i$.  Since this U-duality group is manifest, so are its
subgroups which appear in compactifications on lower dimensional tori. 

It is interesting to discuss the interpretation of spacetime as we go 
from a given space-time 4-torus and its image under an element of 
$SL(5,Z)$. A large space $T^4$ is given by a the (2,0) theory on 
${\tilde T}^4\times S^1$ where the size of the $S^1$ is much smaller 
than any other size of the torus. We then obtain, after a Kaluza-Klein 
reduction, a weakly coupled 4+1 SYM and the weakly coupled Wilson 
lines are the positions of the 0-branes (and hence gravitons) on the 
physical space $T^4$ (i.e., space-time emerges as the moduli
space of vacua of the quantum mechanical system). 

Let us now take the 5-torus and follow a trajectory to an $SL(5,Z)$ 
dual point (which is not included in the $SL(4,Z)$ of the 4+1 SYM torus).
 For convenience, We can do this by enlarging the size of the 5-th 
circle and then shrinking one of the circles of the 4-torus. On the way 
we will pass a point in which all the 5 radii are of approximately 
equal size. At this point we are clearly not justified in performing 
a Kaluza-Klein reduction on any of the circles and we must analyze the 
theory in the full (2,0) theory. We will argue that there is no 
reasonable space-time interpretation at this point.

 It is easy to see
that there is no valid semi-classical moduli space approximation for the 
(2,0) theory.  This can be seen is several complementary ways. Let us 
try and construct the light
degrees of freedom on the moduli space by considering the constant
modes of $B$ (the self dual 2-form in the (2,0) theory).  The 
self-duality condition forces these modes to be time
independent and hence the theory has no light modes and hence no
moduli space of vacua.

In fact, in quantum mechanics (unlike field theory with more than 2
space dimensions) there is never a moduli space of vacua.  Only if the
theory has a parameter, which can be interpreted as $\hbar$, can we
expect an approximate notion of moduli space of vacua.  Then, for
$\hbar=0$ the classical theory can have many static solutions which we
can identify as its moduli space of vacua, $\cal M$.  When $\hbar $ is
small we can study the full quantum theory by restricting the degrees of
freedom to $\cal M$ and quantizing only them.

Returning to our (2,0) theory on $\tilde T^5$, we realize that this theory
does not have a free parameter like $\hbar$.  The six dimensional
theory, because of its self-duality, has fixed $\hbar =1$.  Hence, as
we saw above, it cannot have a semiclassical limit with a moduli space
of vacua.  Instead, we can find a moduli space of vacua by creating an
effective $\hbar$.  One way to do that is to consider the limit of
this theory with $\tilde L_5 \ll \tilde L_{1,2,3,4}$.  Then, by going
through the 4+1 dimensional SYM we find a quantum mechanical system
whose moduli space is $T^4$ with sizes $L_{1,2,3,4} \sim {1 \over
\tilde L_{1,2,3,4}}$.  This interpretation of space-time is not
natural, if we pursue it to the region where all the $\tilde L$'s are
of the same order of magnitude.  As any one of the $\tilde L$'s
is much smaller than the others there is another natural
interpretation of space time.  This is the essence of U-duality in our
construction.  The natural interpretation of the theory in terms of
space time changes as the five radii $\tilde L$ change.

\section{M-theory on K3 and the Heterotic string on $T^3$}

A similar problem occurs when trying to discuss the Matrix model 
description for M-theory on other 4-dimensional manifolds. 
As it is misleading to begin with the SYM prescription, our starting
point will be the (2,0) field theory. 
In this section we discuss
compactifications of M-theory down to seven dimensions, that have 
8 linearly
realized supersymmetries (in the infinite momentum frame). These are
M-theory on $K3$ and the Heterotic string on $T^3$.
We obtain such a theory by compactifying the (2,0) theory on a
5-dimensional base-space that breaks half the supersymmetry. The
manifold that we use is the natural candidate $S^1\times {\tilde {K3}}$ 
(the $\tilde{}$ denotes that it is a different K3 than the space-time 
one). We will
refer to this manifold as the base-space. 

The Matrix model for M-theory on $T^4$ had a space-time interpretation 
only when the base-space ${\tilde T}^5$ degenerated in specific ways. 
We will see that the same is also true here but, unlike in the previous 
case,
in different degenerations of the $S^1\times {\tilde{K3}}$ 
base-space, we obtain
different spacetime interpretations. When the manifold degenerates to
a 4 dimensional base-space in different ways we obtain M-theory on
a large $K3$ and the Heterotic theory on a large $T^3$.  As these 
configurations are
smoothly connected, and we can follow the transition from one limit to
another and the duality between these theories is manifest 
(the technology involved is, in fact, similar to \cite{harvstrom}).

\subsection{M-Theory on $K3$}

We begin by obtaining M-theory on $K3$ in Matrix theory
\cite{arvwilly}. Let us denote the size of the
base-space $S^1$ by $\Sigma_1$ and the volume of the ${\tilde{K3}}$ 
by $V$. The limit of
M-theory on a large $K3$ is obtained by 
\begin{equation}
\begin{array}{rcl}
\Sigma_1, V\rightarrow 0 \\
{\Sigma_1\over V}\ fixed.\\
\end{array}
\label{eq:mlint}
\end{equation}
The second requirement guarantees that the eleven
dimensional Planck scale is fixed. 

When $\Sigma_1$ is much smaller than any length scale of the 
$\tilde{K3}$, the
IR of the base-space theory is approximated by a Kaluza-Klein
reduction on the $S^1$, which is a weakly coupled SYM on
$\tilde{K3}$ (the gauge
coupling of the this SYM is $g^2=\Sigma_1$). By a sort of T-duality, 
which we assume exists\footnote{The precise definition of the T-duality 
for a generic K3 still needs to be clarified (see \cite{ozhori}), we 
will, however, deal later primarily with $T^4/Z_2$ where we can 
explicitly construct it.}. We then obtain a description of the moduli 
space of the theory in terms of 0-branes
moving on a dual $K3$, which is the physical space-time $K3$

In the case of orbifold limits of $\tilde{K3}$, the requirement 
that $S^1$ is
smaller than any length scale of $\tilde{K3}$ cannot be satisfied, 
as there
are 2-cycles of zero size.  This results in the appearance of
additional degrees of freedom in the effective 4+1 SYM. These are
related to the 32 fermions that appear in [24-29]
and will be discussed further in section 3.2.3. For now we restrict
our attention to degrees of freedom that live in the bulk, far away
from any orbifold points.

The remainder of this section will be devoted to a discussion of some 
supporting evidence for this proposal. Let us work in 
the limit
that $\tilde{K3}$ is the orbifold $T^4/Z_2$, whose sizes are
$\Sigma_2,..,\Sigma_5$. 
The relation of the spacetime parameters to the SYM parameters are 
similar to those of M-theory on a
4-torus, and are given in \cite{moshe,fhrs}. These are
\begin{equation}
\begin{array}{rcl}
 {L_i}^2 = {2 \pi R V \over \Sigma_1 {\Sigma_i}^2} \\
{L_p}^6= {R^3 V \over (2 \pi)^3 \Sigma_1} \\
\end{array}
\label{eq:tfour}
\end{equation}
Where $L_i$ (i= 2,...,5) are the spacetime lengths and $L_p$ is the
eleven dimensional Planck length.

In the limit when $g^2 \rightarrow 0$ ($\hbar\rightarrow0$) the
theory becomes semi-classical and the Wilson lines define a moduli
space. This moduli space is interpreted as the classical
compactification manifold in spacetime.  The weakly coupled 4+1 SYM
description of this space is equivalent to the description of 0-branes
moving on this manifold, as can be shown by an explicit T-duality. This
description is valid when this spacetime manifold is much larger than
$L_p$.

Another check that we have identified correctly the Matrix theory is to
reproduce the moduli space of M theory on $K3$. As is often the case
in the infinite momentum frame, modifications to the ground state of
the theory are obtained by modifying the Hamiltonian. In our
case we can modify the base space geometry. The different choices of
base space geometry should give the spacetime moduli space.

We are therefore interested in 5 dimensional manifolds that break
half the supersymmetry. 
There are several discrete choices that need to be made. Within one of 
these choices
we are restricted to metrics on $S^1\times {\tilde{K3}}$ with an $SU(2)$
holonomy. The only parameters of such a metric are the size of the
$S^1$ and a choice of an Einstein (Ricci flat) metric on $\tilde{K3}$. 
The Moduli space is therefore locally $SO(3,19)/(SO(3)\times
SO(19))\times R^+$. This is the correct moduli space of M-theory on $K3$
(and of the Heterotic string on $T^3$) \cite{various}. There are 
actually additional couplings but these are associated the compactness 
of the null directions in the DLCQ description of M-theory on K3, 
and should not be counted as parameters of M-theory on K3. 

The model also has enhanced gauge symmetries at the correct points
of moduli space. If the base-space has certain singularities,
then the T-dual $K3$ has similar singularities. It was shown
in \cite{enhance,arvwilly} that the $N\rightarrow\infty$ Matrix theory 
then
has the additional states that make up the additional gauge bosons.

\medskip
Our main goal is to discuss the duality between M-theory on K3 and the 
Heterotic string on $T^3$\cite{various}. The construction on the 
Heterotic side is more complicated and less complete as the 
degeneration of the base-space is more complicated. Nevertheless we 
can discuss some aspects of it, and suggest how this duality comes 
about in Matrix theory.  

\subsection{Heterotic Theory on $T^3$}

\medskip
\noindent{\it 3.2.1  Heterotic vacuum with $SU(2)^{16}$ Enhanced 
Gauge Symmetry}
\medskip

The case which is easiest to analyze is when $\tilde{K3}=T^4/Z^2$ 
(which has 
16 $A_1$ singularities). This configuration is the one that is most
closely related to the configuration of \cite{tommotl,petr}. Let us
pick one dimension of $T^4/Z_2$, say $\Sigma_5$, and take it to zero, as
well as the volume of the remaining space
$V=\Sigma_1\Sigma_2\Sigma_3\Sigma_4$.  Again, we take $V$ and
$\Sigma_5$ to zero at a fixed ratio.

After the Kaluza-Klein reduction on $\Sigma_5$ we obtain a SYM on 
$S^1\times
(T^3/Z_2)$ with a coupling $g^2=\Sigma_5$ and some boundary conditions
on the gauge fields (We will not discuss these boundary conditions here 
but only state that for $N=1$\footnote{As was explained in 
\cite{enhance}, the state
with N=1 corresponds to a solitonic state in the M-theory. We are now
in position to see how the gauge boson changes from a perturbative
state in the Heterotic string to a solitonic state in M-theory.}, which 
is the only case where we can check them explicitly, they agree with 
\cite{tommotl}
). This model therefore reproduces the Matrix description of the 
Heterotic String on $T^3$ \cite{tommotl,petr}. Note also that the 
instanton
number reverses its sign under the $Z_2$ action, which means that 
these boundary conditions give
 the unique correct extension to the 5-dimensional
manifold $S^1 \times T^4/Z_2$.

We have obtained an Heterotic string theory on $T^3$, and we can write its
parameters in terms of $g^2,\Sigma_{1,2,3,4}$. It is more instructive,
however, to write it in terms of the M-theory on $K3$ parameters 
\ref{eq:tfour}. Doing so,
one obtains
\begin{equation}
\begin{array}{rcl}
T_{string}={L_2L_3L_4L_5\over (2\pi)^5L_p^6} \\
\lambda_7^4={(L_2L_3L_4L_5)^3\over (2\pi)^2L_p^{12}} \\
\end{array}
\label{eq:hetmuda}
\end{equation}
which are the seven dimensional Heterotic/M-theory duality
relations \cite{various}. One can also reproduce more detailed formulas
that relate the radii of the $K3$ to those of the $T^3$\cite{polch}.

\medskip
\noindent{\it 3.2.2. A Conjecture Regarding the $E_8\times E_8$ Case}
\medskip

We are interested in the Heterotic string on $T^3$ in its M-theory
limit. {\it i.e.} when it is M-theory on $S^1/Z_2\times T^3$. We therefore
expect to see a well defined moduli space of this form only when the 
space-time gauge symmetry is $E_8\times E_8$, or a
subgroup of it.

The picture that we suggest is very similar to that of Vafa and
Morrison \cite{vafamorr}. Let us check the case in which the
base-space $\tilde{K3}$ has two $E_8$ singularities. 
The $\tilde{K3}$ can then be written
as an elliptic fibration over $P^1$. On the $P^1$ there are two
singular fibers which contain the $E_8$ singularities and four
additional singularities (where the fiber but not the $\tilde{K3}$
degenerates). We are interested in the limit in which a pair of the
additional singularities approach each $E_8$ loci. In that case the
base becomes a thin long cylinder capped in the vicinity of the $E_8$
singularity. Throughout the cylinder, as long as we are away from the
singularities, the fiber has a constant complex structure
parameter. 

 We are interested in reducing the (2,0) theory on the small circle
which is a part of the cylindrical base of the elliptic fibration. We
take therefore all the other dimensions of the base space to be of the
same order of magnitude, and larger than this small circle. Note that
the size of the fiber is a parameter of the theory (unlike in
F-theory). In this configuration the (2,0) theory has a mass gap and
we can perform a Kaluza-Klein reduction on the small circle.  The base
space of the resulting 4+1 SYM looks like $S^1\times T^2\times I$
where the first $S^1$ is outside $\tilde{K3}$,
 the $T^2$ is the fiber and
$I$ is an interval, which is what is left from the cylinder after the
Kaluza-Klein reduction.  On this space we have a weakly coupled gauge
theory with some boundary conditions on the gauge fields and matter
fields. We can now have four Wilson lines on this space which give us
the four compact space coordinates. Unfortunately, we do not know how 
to calculate the boundary conditions in this
picture, so we can not verify that the moduli space of Wilson lines is 
indeed $T^3\times (S^1/Z_2)$. 

One more comment is in order. When we deform away from the $E_8\times 
E_8$ loci, the degeneration of $\tilde{K3}$ is generically very 
complicated. 
There is no guarantee that for such a degeneration there will be any 
sensible description of the low energy in terms of any 4+1 SYM on a 
manifold. This is different from other approaches taken to the 
Heterotic Matrix theory in [14-19].   

\medskip
\noindent{\it 3.2.3 Additional Degrees of Freedom}
\medskip

Another important difference is the way that the fermions in the 
Heterotic Matrix model are
treated. In our picture the fermions are to be understood as
fermionization of the bosons $\int B$ over shrunken cycles. As such
they are localized at the singularities and are not allowed to
move. Enhanced symmetry is obtained in a geometric way in which the
mixing of the compact space parameters and the $E_8\times E_8$ Wilson
lines is apparent.

More important is the fact that we do not need to add these fermions
by hand \cite{evashumit}. They are automatically provided by the
(2,0) definition of the theory. At no point of the discussion do we
 need to take a circle, in the Matrix description of M-theory on $T^4$,
orbifold it and add 8-branes. Rather the 8-branes are generated in the
effective space-time by the existence of additional degrees of freedom
in a specific degeneration of the (2,0) base-space.  We know that
there are 8-branes only through the existence of the 32 fermions. When
we calculate any low energy scattering the fermions contribute to the
scattering amplitude such that a low energy observer interprets the
result as the existence of 8-branes.

\subsection{Duality}

To summarize both M on $K3$ 
and the Heterotic string on $T^3$ are described by the same
model. Hence, duality is manifest. In one limit of the geometry of the 
base-space we obtain the a
description of the weakly coupled low-energy of the Heterotic string
and in another limit that of M theory on $K3$. The transition between
these two limits, as expected, goes through a region in which
the compact part of space-time is not well defined.

\centerline{\bf Acknowledgments}

The results presented in this talk were obtained in collaboration with 
M. Rozali and N. Seiberg. Both myself and my collaborators would like 
to thank O. Aharony, P. Aspinwall, T. Banks,
D. Berenstein, R. Corrado, J. Distler, M. Douglas, R. Entin, D. Finnel,
W. Fischler, O. Ganor, P. Horava, S. Kachru, S. Sethi,
S. Shenker and E. Witten for useful and illuminating discussions. 
This work was
supported by NSF grant NSF PHY-9513835.

\end{document}